# Theoretical kinetic study of the low temperature oxidation of ethanol

R. Fournet[1*], P.A. Glaude[1], R. Bounaceur[1], M. Molière[2]

[1]Département de Chimie Physique des Réactions,
CNRS-Nancy University, France ENSIC, 1 rue Grandville, BP 20451, 54001 Nancy Cedex, France

[2]GE Energy Product-Europe,
20 avenue du Maréchal Juin, BP 379, 90007 Belfort, France

**Abstract**
In order to improve the understanding of the low temperature combustion of ethanol, high-level *ab initio* calculations were performed for elementary reactions involving hydroxyethylperoxy radicals. These radicals come from the addition of hydroxethyl radicals (•$CH_3CHOH$ and •$CH_2CH_2OH$) on oxygen molecule. Unimolecular reactions involving hydroxyethylperoxy radicals and their radical products were studied at the CBS-QB3 level of theory. The results allowed to highlight the principal ways of decomposition of these radicals. Calculations of potential energy surfaces showed that the principal channels lead to the formation of $HO_2$ radicals which can be considered, at low temperature, as slightly reactive. However, in the case of $CH_3CH(OOH)O•$ radicals, a route of decomposition yields H atom and formic peracid, which is a branching agent that can strongly enhance the reactivity of ethanol in low temperature oxidation. In addition to these analyses, high-pressure limit rate constants were derived in the temperature range 400 to 1000 K.

**Introduction**

Many researches in the field of energy are currently carried out in order to limit $CO_2$ emissions. Among alternative fuels to natural gas, gasolines or diesel fuels in engines or combustion devices, ethanol can find an increasing place in the next years. However, if the studies concerning the combustion of methanol are rather numerous, the researches performed on the combustion of ethanol are scarce. In particular, the low temperature pathways can be of importance in the auto-ignition process in engine or for process security in case of a leak of ethanol close to hot surfaces. A few numbers of theoretical studies, involving reactions of hydroxyethyl radicals with oxygen, can be found in the literature. Da Silva et al. calculated the chemically activated hydroxyethyl + $O_2$ systems. Olivella and Solé[1] studied from *ab initio* calculations the unimolecular decomposition of •$OOCH_2CH_2OH$ radical. Each pathway has been characterized by means of density functional theory (B3LYP) and quantum-mechanical (CCSD(T)) calculations with basis sets ranging from the 6-31G(d,p) to the 6-311+G(3df,2p). They found that the preferred pathway corresponds to the 1,5-hydrogen shift. Kuwata et al.[2] used composite methods, RRKM / master equation and VTST to calculate the rate constants for the 1,5-hydrogen shift for the radical •$OOCH_2CH_2OH$. They found a barrier of 23.59 kcal/mol for this reaction. More recently, Zador et al.[3] investigated the reaction of hydroxyethyl radicals with $O_2$ by means of theoretical calculations and experimental study. They performed high-level *ab initio* calculations of the potential energy surfaces coupled with master equation methods to compute rate coefficients for the reactions:

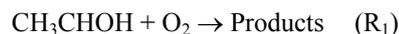
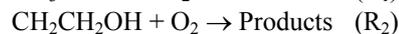

$CH_3CHOH + O_2 \rightarrow$ Products    ($R_1$)
$CH_2CH_2OH + O_2 \rightarrow$ Products    ($R_2$)

They found that the major products of reaction $R_1$ are acetaldehyde and $HO_2$, while for reaction 2, several products can be formed, such as 2 HCHO and OH, or vinyl alcohol and $HO_2$.

In this studied, we performed theoretical calculations at the CBS-QB3 level of theory in order to highlight the decomposition channels of hydroxyl peroxyethyls radicals (•$OOCH_2CH_2OH$ and $CH_3CH(OO•)OH$). From these calculations, we derived rate parameters for important reactions.

**Computational method**

Calculations were performed with Gaussian 03 Rev. C.02[4]. The composite method CBS-QB3[5] was applied for all stationary geometries and transition states involved in the reaction schemes. Vibrational frequencies calculated at the B3LYP/cbsb7 level of theory confirm that all transition states (TS) have exactly one imaginary frequency. IRC calculations have been done at the same level of theory to ensure that the transition states connect correctly the reactants to the products Thermochemical data were derived from CBS-QB3 energy and frequencies. Enthalpies of formation ($\Delta H_f°$) of species involved in this study have been calculated using isodesmic

---

reactions, excepted for species (molecules) for which more accurate experimental enthalpies of formation [Nist] can be found in the literature. Thanks to the conservation of the total number and types of bonds, very good results can be obtained by the cancellation of errors on the two sides of the reaction. Several isodesmic reactions have been considered for the calculation of $\Delta H_f°$ in order to obtain an average value.

Internal rotors have been taken into account by performing a scan along the rotation at the B3LYP 6-31+G(d,p) level of theory, in order to estimate the energy barrier of each hindered rotor.

Spin contamination was observed for radicals and transition states at the CBS-QB3 level of theory (0.8 < <s$^2$> <1). However, it must be noted that, in the CBS-QB3 method, an empirical correction for spin contamination is performed for the energy calculation.

The high-pressure rate constants involved in the mechanisms were calculated with the software CHEMRATE[6], using transition state theory (TST):

$$k_{uni} = rpd\, \kappa(T) \frac{k_b T}{h} \exp\left(\frac{\Delta S^{\neq}}{R}\right) \exp\left(-\frac{\Delta H^{\neq}}{RT}\right) \quad (1)$$

where $\Delta S^{\neq}$ and $\Delta H^{\neq}$ are, respectively, the entropy and enthalpy of activation and *rpd* is the reaction path degeneracy. The activation enthalpies involved in TST theory were calculated by taking into account the enthalpies of reaction calculated with isodesmic reactions. By example, in an elementary unimolecular reaction (reactant (R) → products (P)) we can write:

$$\Delta H^{\neq}(R \to P) = \left(\Delta H^{\neq}_{1(CBS-QB3)} + \Delta H^{\neq}_{-1(CBS-QB3)} + \Delta H_r \text{ (isodesmic)}\right)/2 \quad (2)$$

and

$$\Delta H^{\neq}(P \to R) = \left(\Delta H^{\neq}_{1(CBS-QB3)} + \Delta H^{\neq}_{-1(CBS-QB3)} - \Delta H_r \text{ (isodesmic)}\right)/2 \quad (3)$$

where $\Delta H^{\neq}_{1(CBS-QB3)}$ and $\Delta H^{\neq}_{-1(CBS-QB3)}$ are, respectively, the enthalpy of activation for the forward and back reactions calculated at a temperature T(K). $\Delta H_{r(isodesmic)}$ corresponds to the enthalpy of reaction calculated at T(K) using NASA polynomial obtained from calculated isodesmic enthalpies of formation and entropy at 298 K, and heat capacities of reactants and products presented in Table 1.

For reactions involving H-transfer, a transmission coefficient, namely $\kappa(T)$, has been calculated in order to take into account tunneling effect. We used an approximation to $\kappa$ provided by Skodje and Truhlar :

for $\beta \leq \alpha$

$$\kappa(T) = \frac{\beta \pi / \alpha}{\sin(\beta \pi / \alpha)} - \frac{\beta}{\alpha - \beta} e^{[(\beta - \alpha)(\Delta V^{\neq} - V)]} \quad (4)$$

for $\alpha \leq \beta$

$$\kappa(T) = \frac{\beta}{\beta - \alpha} \left[e^{[(\beta - \alpha)(\Delta V^{\neq} - V)]} - 1\right]$$

where $\alpha = \frac{2\pi}{h\, \text{Im}(\nu^{\neq})}, \beta = \frac{1}{k_B T}$

In equation (4), $\nu^{\neq}$ is the imaginary frequency associated with the reaction coordinate. $\Delta V^{\neq}$ is the zero-point-including potential energy difference between the TS structure and the reactants, and V is 0 for an exoergic reaction and the zero-point-including potential energy difference between reactants and products for an endoergic reaction. As mentioned by authors, an inspection of the power series expansion for the exponential in these equations indicates that neither expression diverges as α and β become arbitrarily close to equal.

The kinetic parameters were obtained by fitting the rate constant values obtained from TST at several temperatures between 400 and 1000 K with:

$$k_{\infty} = A\, T^n \exp(-E/RT) \quad (5)$$

where A, n, E are the parameters of the modified Arrhenius equation and $k_{\infty}$ is the high–pressure limit rate constant.

**Thermochemical data**

Thermochemical data ($\Delta_f H°$, $S°$, $C_p°$) for all the free species involved in this study were derived from energy and frequency calculations and are collected in Table 1. In the CBS-QB3 method, harmonic frequencies, calculated at the B3LYP/cbsb7 level of theory, were scaled by a factor 0.99. Accurate experimental enthalpies of formation found in the literature were used instead of those calculated. Thus, in Table 1, enthalpies of formation written in italic corresponds to experimental data found in the NIST Chemical Database[7].

**Theoritical calculations**
*A – Decomposition of •OOCH$_2$CH$_2$OH*

Figure 1 presents the ZPE corrected potential energy surfaces for the decomposition of •OOCH$_2$CH$_2$OH. As shown in figure 1, isomerization yielding to HOOCH$_2$CH$_2$O represents the easiest way of decomposition of this hydroxyethylperoxy radical, with an energy barrier of 21.1 kcal mol$^{-1}$ at 0K.

*Table 1: thermochemical data for species involved in decomposition reactions of hydroxyethylperoxy radicals*
*$\Delta_f H°(298K)$ is expressed in kcal.mol$^{-1}$, $S°(298 K)$ in cal.mol$^{-1}$ and $C_p°(T)$ in cal.mol$^{-1}$.K$^{-1}$*

| Species | $\Delta_f H°$(298K) | $S°$(298 K) | $C_p°(T)$ | | | | | |
|---|---|---|---|---|---|---|---|---|
| | | | 300 | 500 | 800 | 1100 | 1200 | 1500 |
| $CH_3^{(7)}$ | 34.82 | 46.59 | 9.43 | 10.89 | 12.87 | 14.04 | 15.04 | 16.22 |
| $HO_2^{(7)}$ | 2.94 | 54.71 | 8.30 | 9.39 | 10.68 | 11.27 | 11.74 | 12.26 |
| $OH^{(7)}$ | 9.32 | 43.88 | 7.15 | 7.04 | 7.10 | 7.26 | 7.45 | 7.74 |
| $C_2H_3OH^{(7)}$ | -30.60 | 62.89 | 14.39 | 20.01 | 25.12 | 27.36 | 29.10 | 31.02 |
| $CH_3COOH^{(7)}$ | -103.50 | 68.97 | 16.22 | 22.55 | 29.08 | 32.03 | 34.25 | 36.59 |
| $HCOO_2H$ | -66.67 | 68.02 | 15.65 | 20.00 | 24.28 | 26.11 | 27.41 | 28.67 |
| (epoxide–OH) | -57.70 | 66.40 | 15.99 | 23.42 | 30.00 | 32.68 | 34.65 | 36.76 |
| $OOCH_2CH_2OH$ | -39.28 | 81.98 | 22.09 | 30.30 | 38.36 | 41.73 | 44.16 | 46.63 |
| $CHOCH_2OH$ | -72.84 | 71.40 | 17.53 | 23.91 | 29.87 | 32.38 | 34.24 | 36.22 |
| $CH_3CH(OO)OH$ | -51.40 | 79.71 | 22.73 | 31.27 | 38.97 | 42.03 | 44.22 | 46.50 |
| $HOOCH_2CHOH$ | -31.94 | 83.81 | 24.62 | 32.93 | 40.73 | 43.78 | 45.85 | 47.84 |
| $HOOCH_2CH_2O$ | -20.02 | 80.60 | 22.59 | 31.44 | 40.03 | 43.53 | 45.97 | 48.35 |
| $CH_2CH(OOH)OH$ | -32.62 | 81.44 | 25.49 | 33.28 | 40.19 | 43.07 | 45.17 | 47.35 |
| $CH_3CHO(OOH)$ | -33.26 | 78.02 | 22.85 | 31.21 | 39.13 | 42.54 | 45.04 | 47.65 |

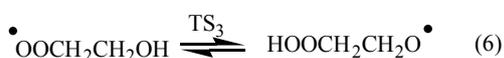

$$\bullet OOCH_2CH_2OH \overset{TS_3}{\rightleftharpoons} HOOCH_2CH_2O\bullet \quad (6)$$

The energy barrier is relatively low since the isomerization involved a six-member ring in the transition state (TS$_3$) with a very low ring strain energy. The HOOCH$_2$CH$_2$O• radical reacts easily by β-scission of the C-C bond to produce HCHO and HOOCH$_2$• which yields afterwards formaldehyde and OH radical according to:

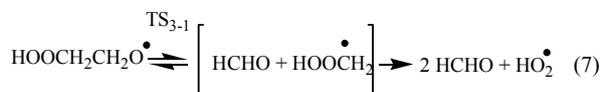

$$HOOCH_2CH_2O\bullet \overset{TS_{3-1}}{\rightleftharpoons} [HCHO + HOOCH_2\bullet] \rightarrow 2\ HCHO + HO_2\bullet \quad (7)$$

The energy barrier involved in the reaction (7) is equal to 4.8 kcal mol$^{-1}$ and shows the weakness of the C-C bond in the vicinity of two oxygenated groups.

Energy barrier values for isomerizations via transition states TS$_{1D}$ and TS$_2$ are, respectively, 18.8 and 5.5 kcal mol$^{-1}$ higher than that calculated for reaction (6). It can be explained by the number of atoms involved in the H transfer during the reactions. Indeed, for TS$_{1D}$ and TS$_2$, respectively, four and five-member rings are involved. It must be noted that for reaction implying TS$_{1D}$, the formation of a hydroperoxide bond located in β position of the radical center leads directly to the formation of acetaldehyde and OH radical :

$$\bullet OOCH_2CH_2OH \overset{TS_{1D}}{\rightleftharpoons} [HOO\bullet CHCH_2OH] \rightarrow \bullet OH + OH-CH_2-CHO \quad (8)$$

HOOCH$_2$CHOH radical formed by isomerization via a five-member ring (TS$_2$) can react by two ways according to reactions 9 and 10:

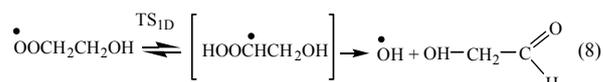

$$HOOCH_2\bullet CHOH \overset{TS_{2-1}}{\longrightarrow} C_2H_3OH + HO_2\bullet \quad (9)$$

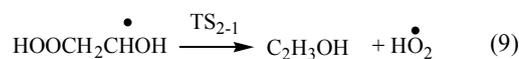
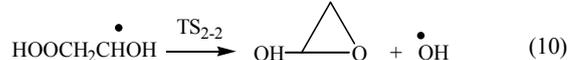

$$HOOCH_2\bullet CHOH \overset{TS_{2-2}}{\longrightarrow} OH-\triangle-O + \bullet OH \quad (10)$$

The barrier heights for reactions (9) and (10) are close and equal, respectively, 13.9 and 12.4 kcal mol$^{-1}$.

The last reaction of •OOCH$_2$CH$_2$OH is a direct decomposition to C$_2$H$_3$OH and HO$_2$ (figure 1). From TS$_{2D}$, it can be seen that the reaction involved simultaneously a H transfer and the breaking of the C-OOH bond, as shown in figure 2. However the energy barrier is high (30.8 kcal mol$^{-1}$) and cannot play a role in the oxidation of ethanol at low temperature.

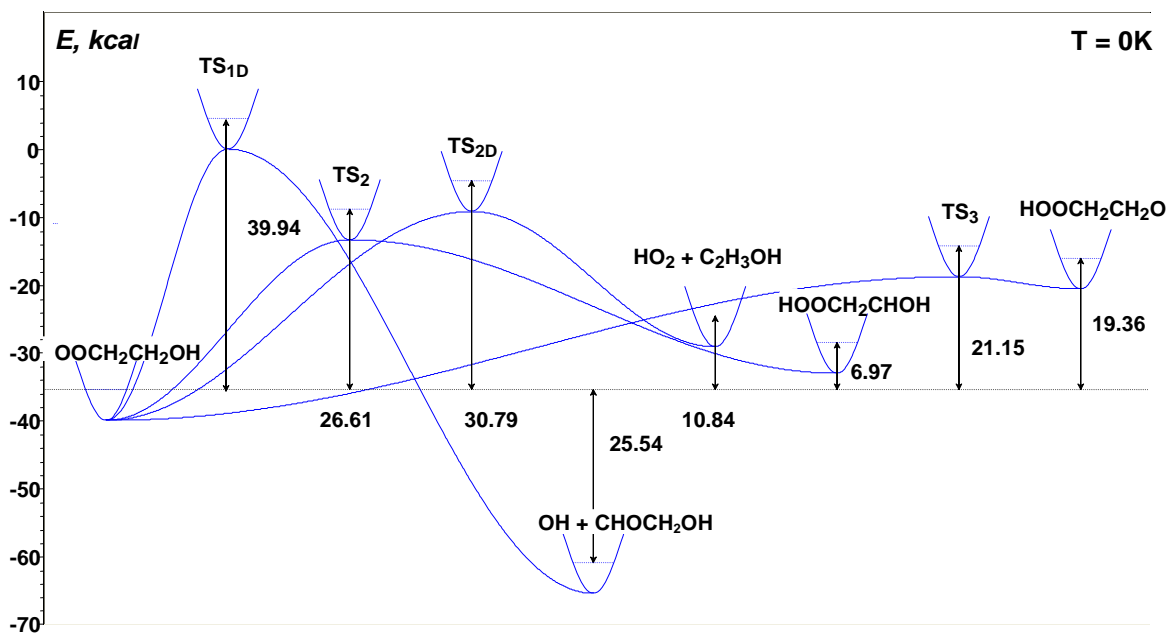

*Figure 1: ZPE corrected potential energy surfaces for the decomposition of •OOCH$_2$CH$_2$OH. The energies were obtained at the CBS-QB3 level of theory.*

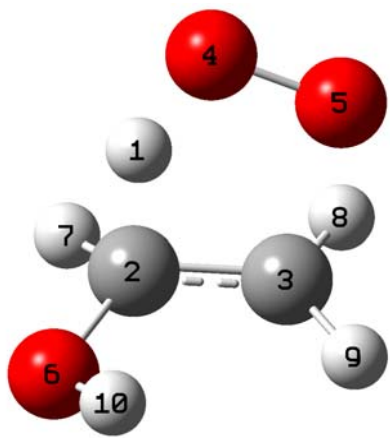

*Figure 2: Transition state TS$_{2D}$ involved in the direct decomposition of OOCH$_2$CH$_2$OH. Geometry obtains at the B3LYP/CBS7 level of theory.*

In summary the easiest channel for the decomposition of •OOCH$_2$CH$_2$OH is the isomerization by H transfer from hydroxyl group leading to the formation of hydroperoxide radical, which in its turn reacts by β-scission, yielding OH and two formaldehyde molecules. The other routes involve higher activation energies and should not have any important kinetic influence during the oxidation of ethanol at low temperature.

*B – Decomposition of CH$_3$CH(OO•)OH*

Figure 3 shows the five routes of decomposition of CH$_3$CH(OO•)OH found at the CBS-QB3 level of theory. The easiest reaction involves a direct decomposition according to:

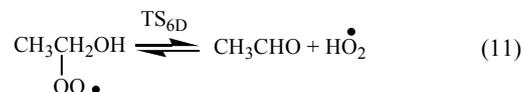

(11)

As mentioned by Zador et al.[3], the product corresponds, in fact, to a complex in which remains a weak H-bond as shown in figure 4. The length of the hydrogen bond (C$_3$ and H$_{10}$ in figure 4) is 1.754 Å. This reaction is close to the Waddington mechanism[8] which corresponds to H transfer from the OH group with a simultaneous breaking of the C-OO bond. This reaction is favored and the energy barrier is very low (11.3 kcal/mol). The formation of HO$_2$ from reaction 11, will induce a decrease of the reactivity of ethanol since HO$_2$ reacts mainly by termination in the low temperature range.

The second channel corresponds to an isomerization of CH$_3$CH(OO•)OH by a H transfer from the OH group, via a 6-member ring and according to reaction 12:

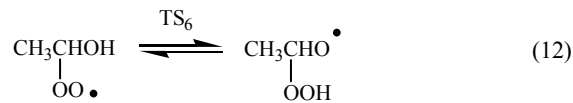

(12)

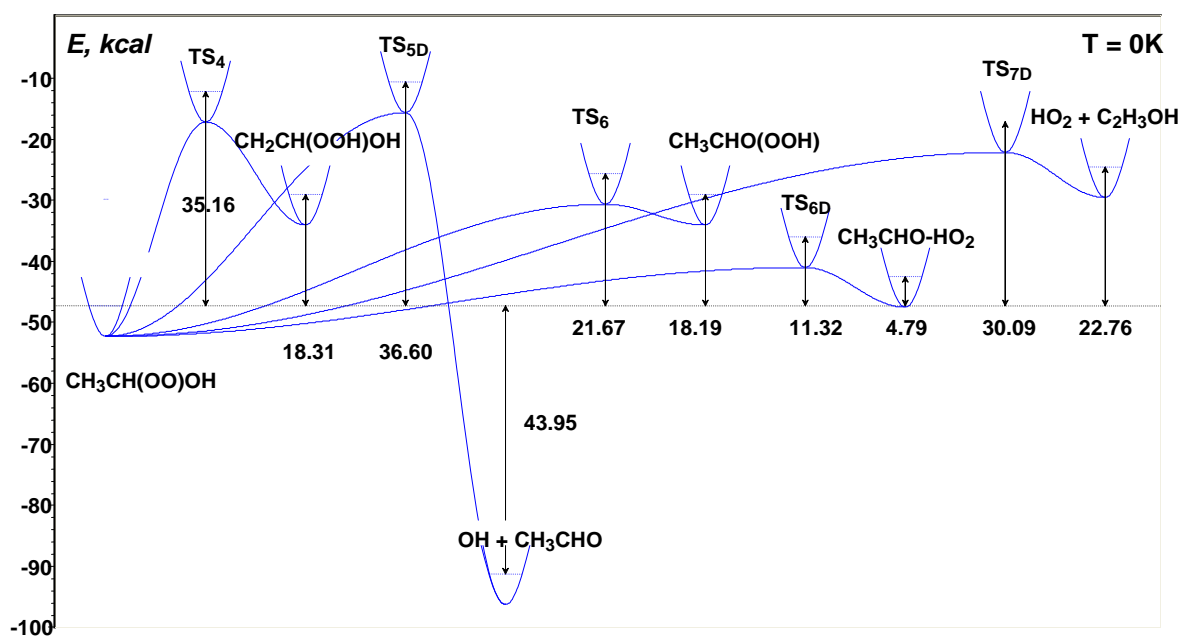

Figure 3: *ZPE corrected potential energy surfaces for the decomposition of CH$_3$CH(OO•)OH. The energies were obtained at the CBS-QB3 level of theory.*

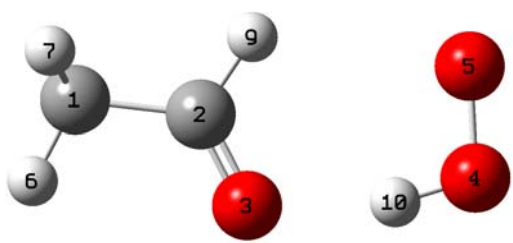

Figure 4: *Complex formed in the direct decomposition of CH$_3$CH(OO•)OH (reaction 11). Geometry obtained at the B3LYP/CBS7 level of theory.*

The barrier height is 10.4 kcal mol$^{-1}$ higher than that involves in reaction 11 and corresponds to a factor 1000 at 750K between the two rate constants.

CH$_3$CH(OO•)OH can decompose by three other channels according to reactions:

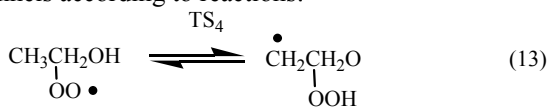  (13)

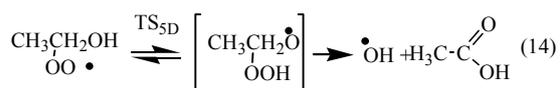  (14)

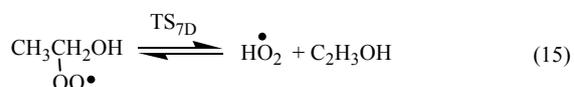  (15)

These three reactions correspond to very high energy barriers comparing to reaction 11 (between 18.8 and 25.3 kcal mol$^{-1}$ higher) and should not have any kinetic influence in the combustion of ethanol at low temperature

Even if reaction (11) represents the easiest route of decomposition of CH$_3$CH(OO•)OH, we studied the decomposition reactions of the product formed in reaction 12 (CH$_3$CH(OOH)O•) and we highlighted three possible channels as shown in figure 5. Among the three routes, one yields to the formation of HO$_2$ (reaction 16) while the other (reaction 17) leads to the formation of a peracid (HCOO$_2$H) with represents a branching agent and can increase the reactivity of ethanol:

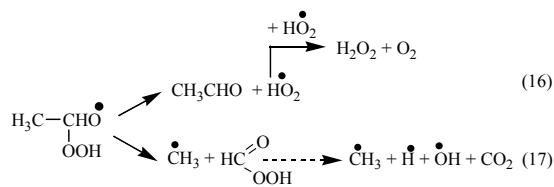 (16) (17)

The energy barrier of reaction 17 is slightly lower than that involved in the formation of HO$_2$ (10.3 kcal mol$^{-1}$ against 13 kcal mol$^{-1}$).

*C – Rate parameters*

Table 2 summarizes the rate constants of important reactions involved in the decomposition of hydroxyethylperoxy radicals and expressed from equation 5. The rate parameters have been calculated between 400 and 1000 K, i.e. in the low temperature range. The parameters in Table 2 correspond to high-pressure limit rate constants.

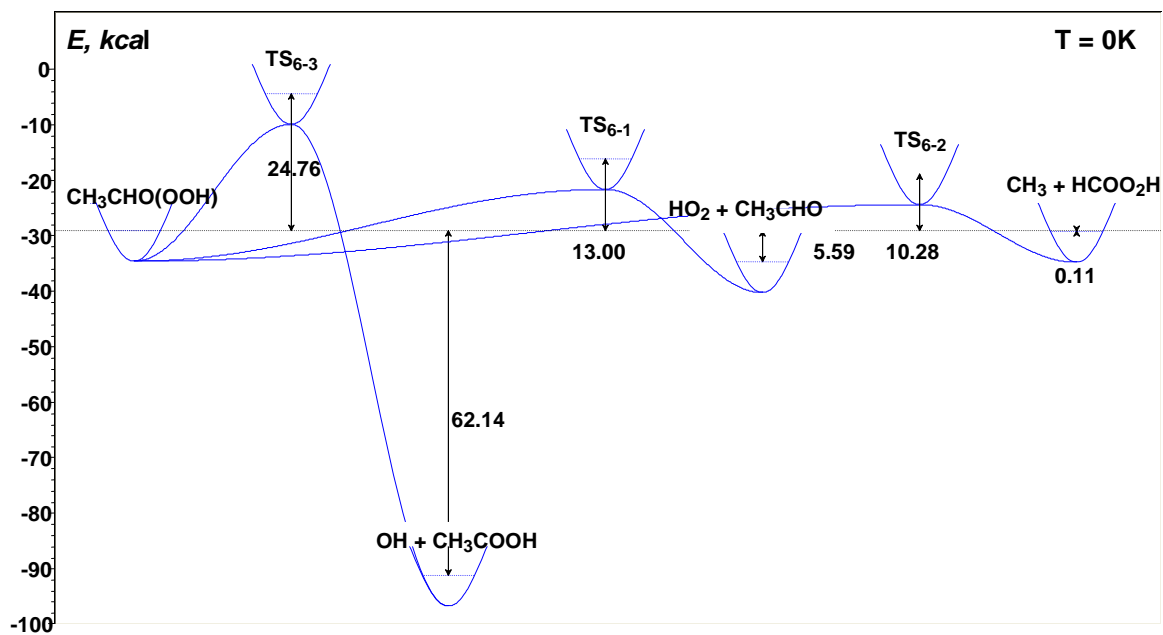

*Figure 5*: ZPE corrected potential energy surfaces for the decomposition of $(CH_3CH(OOH)O\bullet)$. The energies were obtained at the CBS-QB3 level of theory

*Table 2*: rate parameters calculated at the CBS-QB3 level of theory for important reactions involved in the low temperature combustion of ethanol

| Reaction | A (s$^{-1}$) | n | E (kcal/mol) |
|---|---|---|---|
| 6 (TS$_3$) | 2.84 10$^8$ | 0.736 | 19.20 |
| 11 (TS$_{6D}$) | 1.69 10$^{11}$ | 0.342 | 10.56 |
| 12 (TS$_6$) | 5.98 10$^{11}$ | 0.017 | 21.56 |
| 16 (TS$_{6-1}$) | 2.43 10$^{12}$ | 0.365 | 13.30 |
| 17 (TS$_{6-2}$) | 5.23 10$^{11}$ | 0.690 | 10.71 |

**Conclusion**

This paper presents a theoretical study of the oxidation of ethanol at low temperature by means of quantum calculations. Decompositions of hydroxyethylperoxy radicals, formed by addition of hydroxyethyl radicals with O$_2$, have been investigated at the CBS-QB3 level of theory. For each hydroxyethylperoxy radicals a major route of decomposition exists. •OOCH$_2$CH$_2$OH reacts mainly by internal H transfer from the OH group, *via* a 6-member ring and yields, finally, formaldehyde and OH radical. CH$_3$CH(OO•)OH reacts by direct decomposition leading to the formation of H-bonded complex CH$_3$CHO...HO$_2$ which can easily decompose to acetaldehyde and HO$_2$. A minor way of decomposition of CH$_3$CH(OO•)OH yields CH$_3$CH(OO)O• by isomerization. β-scissions of this last radical lead either to the formation of a peracid, which represents a branching agent, or to the formation of acetaldehyde and HO$_2$, which react slightly at low temperature**.**

**Acknowledgement**

This work has been supported by General Electric Energy Products – Europe.